%% file: apssamp.tex
\documentclass[
 reprint,
superscriptaddress,
 amsmath,amssymb,
 aps
]{revtex4-2}
\usepackage{appendix}
\usepackage{graphicx}
\usepackage[dvipsnames]{xcolor}
\usepackage{float}
\usepackage{dcolumn}
\usepackage{bm}

\input{commands}

\begin{document}
\preprint{APS/123-QED}

\title{Interactions and Reconnections of Four-Dimensional Quantum Vortices}
\author{H. A. J. Middleton-Spencer}
\email{h.a.j.middleton-spencer@bham.ac.uk}
\affiliation{School of Physics and Astronomy, University of Birmingham, UK}
\author{B. McCanna}
\affiliation{School of Computer Science, University of Birmingham, UK}
\author{D. Proment}
\affiliation{School of Engineering, Mathematics and Physics, University of East Anglia, UK}
\author{H. M. Price}
\affiliation{School of Physics and Astronomy, University of Birmingham, UK}

\date{\today}
\begin{abstract}
\noindent Interactions and reconnections of vortices are fundamental in many areas of physics, including classical and quantum fluids where they are central to understanding phenomena such as turbulence. In three-dimensional (3D) superfluids, quantum vortices are one-dimensional (1D) filaments that can intersect, reconnect, and recoil with irreversible dynamics described by near-universal scaling laws. We explore quantum-vortex reconnections in a four-dimensional (4D) superfluid, where a vortex is a two-dimensional (2D) surface. Using real-time numerical simulations, we find much richer behaviour than in 3D, including stable intersecting vortex surfaces which do not reconnect, and unusual vortex reconnections which occur without significant energy dissipation, which may indicate quasi-reversible dynamics. Our work raises many interesting questions about vortex physics in extra-dimensional systems, and may be further extended to look at vortex dynamics and turbulence in higher-dimensional spaces.
\end{abstract}

\maketitle

\section{Introduction}
Vortex reconnections have long been of fundamental importance in many fields, ranging from plasma physics~\cite{che2011current,RevModPhys.82.603} to classical~\cite{kida1994vortex,pumir1987numerical,kleckner2013creation} and quantum fluids~\cite{bewley2008characterization,paoletti2010reconnection,serafini2017, fonda2014direct, nazarenko2003analytical}. 
These processes play a vital role in fluid dynamics as they provide a mechanism for the distribution of energy to multiple length scales throughout the system, and so underlie phenomena such as turbulence~\cite{kida1994vortex, baggaley2012vortex, madeira2020quantum,middleton2023strong,barenghi2023types,PhysRevE.93.061103}, energy dissipation and relaxation towards equilibrium~\cite{leadbeater2001sound, villois2020irreversible, xhani2020critical}. 

In quantum fluids, vortex reconnections have been observed experimentally both in liquid Helium~\cite{bewley2008characterization,paoletti2010reconnection} and in Bose-Einstein condensates (BECs)~\cite{serafini2017}. 
Unlike the arbitrary circulation and size allowed in classical fluids, quantum vortices are characterised by a quantized circulation around a fixed-size ``vortex core''  of vanishing superfluid density~\cite{vinen1961detection,PhysRevLett.43.214,pitaevskii2003, pethick2008, cooper2008rapidly, fetter2009, madison2000, madison2001, matthews1999,abo2001observation,verhelst2017vortex}. In 3D, a quantum vortex has a core described by a 1D tube or ``filament'', corresponding to a topological defect in the order parameter. 
Vortex reconnections are then dramatic events, occurring at a specific time, $t_0$, when the core topology changes~\cite{Schwarz1985, Koplik1993, Alamri2008,nazarenko2003analytical}. Physically this corresponds to, for example, two vortex filaments intersecting and exchanging tails, before moving apart.

Remarkably, the dynamics of 3D vortex reconnections is universal at small scales~\cite{villois2017universal}, with the minimum distance between the vortex lines, $\delta$, varying near the reconnection as
\begin{equation}
    \delta(t) \approx \mathcal{A}^{\pm} (\kappa |t-t_0|)^{1/2},
    \label{eq:reconnection}
\end{equation}
where $\kappa=h/m$ is the quantum of circulation, with Planck's constant $h$ and particle mass $m$, and where $\mathcal{A}^{-}$ ($\mathcal{A}^{+}$) is a dimensionless scaling parameter before (after) the reconnection time, $t_0$~\cite{nazarenko2003analytical, galantucci2019crossover, enciso2021approximation}. 
Generically, $\mathcal{A}^+>\mathcal{A}^-$, reflecting that vortices repel faster than they approach \cite{zuccher2012quantum, PhysRevFluids.5.104701,villois2020irreversible,galantucci2019crossover}; this is linked to the transfer of energy from the vortex core into sound and Kelvin waves~\cite{svistunov1995superfluid, PhysRevFluids.5.104701,villois2020irreversible} as well as a loss of vortex-core length, meaning that 3D vortex reconnections are statistically irreversible. 

In this paper, we investigate vortex interactions and reconnections in a superfluid with {\it four spatial dimensions}. 
This builds on recent work studying how, in a 4D BEC, quantum vortices take the form of 2D planes~\cite{mccanna2021superfluid}, or more generally surfaces~\cite{mccanna2023curved,mccanna2023curved2}. 
As we now show with real-time dynamical simulations, the resulting 4D quantum vortex reconnections are much richer than their 3D counterpart, with very different dynamics being observed depending on the initial reconnecting surface. We focus on three classes of configurations for which we observe, respectively, no reconnections; 3D-like reconnections accompanied by the transfer of energy into sound waves; and unusual 4D reconnections, in which there is no appreciable energy transfer, suggesting the dynamics may be quasi-reversible in a statistical sense. Unlike for extended vortex lines in 3D, the vortex in each case remains a single connected surface throughout the real-time dynamics, even as the topology of this surface changes through reconnection. 

{We present here a minimal model for a 4D BEC. In the future, it will be interesting to explore how one can form at 4D superfluid by the implementation of so-called synthetic dimensions~\cite{Ozawa2019} Furthermore, it could also be insightful draw connections with other higher-dimensional mathematical models describing high-energy/string theory physical systems~\cite{hashimoto2005reconnection,hanany2005reconnection}.
\begin{figure*}
    \centering
    {\includegraphics[trim={4cm 3.5cm 4cm 5.5cm},clip,width=0.9\textwidth]{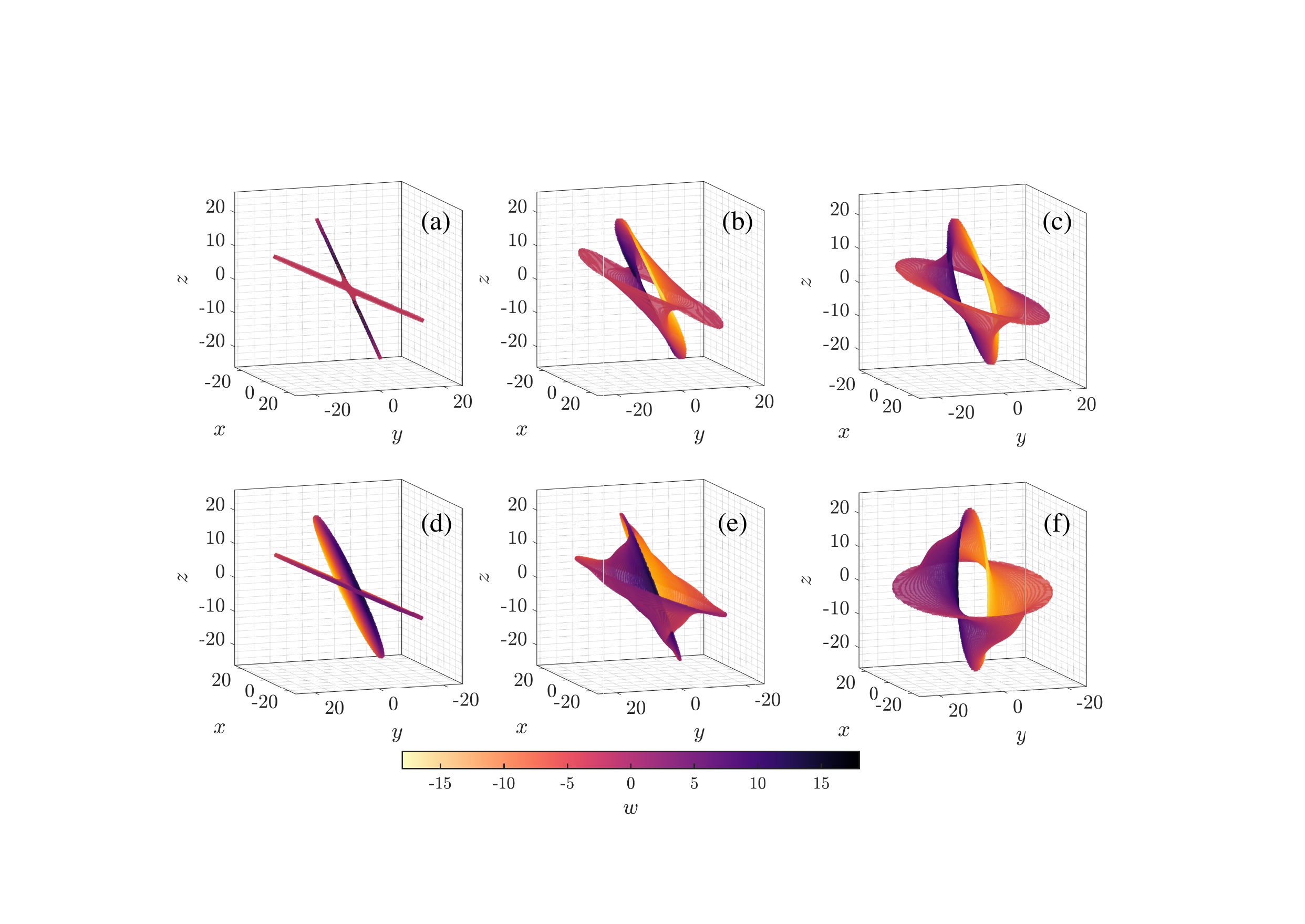}}
    \caption{Numerical real-time dynamics of the 4D GPE for an input wave function containing two vortices intersecting at the origin [c.f. Eq.~\eqref{eq:input}].  (\textit{Top Row:}) The input state has $\theta_1= 50^\circ$, $\theta_2=0$ [c.f. Eq.~\eqref{eq:skewMatrix}], corresponding to planar-aligned vortices, shown for (a) $t=0$, (b) $t=50$, (c) $t=100$, which is classified as a 4D generalisation of 3D reconnection. (\textit{Bottom Row:}) The input state has $\theta_1= -\theta_2= 50^\circ$, corresponding to anti-aligned vortices, shown for (d) $t=0$, (e) $t=50$, (f) $t=100$, which is a new type of reconnection not observed in 3D systems. In both cases the vortices reconnect, opening up a hole around the origin which grows in time. The figures are presented in the $(x,y,z)$ subspace, with $w$ denoted by colour.
    }
    \label{fig:Fig_1}
\end{figure*}

The structure of this paper is as follows: in Section~\ref{sec:back} we  review 4D quantum vortices, defining the three classes of configurations to be studied, before outlining the numerical methods that we use. In Section~\ref{sec:results}, we present our numerical results, focusing on the physics of each class of studied configurations in turn. Finally, in Section~\ref{sec:concl} we discuss our results and draw conclusions. 

\section{Background and Methods} \label{sec:back}

\subsection{4D Quantum Vortices} 
We model our system as a bosonic quantum fluid in the zero-temperature mean-field limit, where the condensate is described by a single order parameter wave function, $\psi$. 
The dynamics of $\psi$ are governed by the dimensionless 4D Gross-Pitaevskii equation (GPE)~\cite{mccanna2021superfluid,mccanna2023curved,mccanna2023curved2}
\begin{equation}
i\frac{\partial\psi'(\mathbf{x}',t')}{\partial t'} = \Big[-\frac{1}{2}{\nabla'}^2+V(\mathbf{x}')+|\psi'(\mathbf{x'},t')|^2\Big]\psi'(\mathbf{x'},t'), \label{eq:4DGPE}
\end{equation}
where $\mathbf{x}'=(x',y',z',w')$, denoting the four Cartesian dimensions. Equation (\ref{eq:4DGPE}) has been rescaled into natural units: $\mathbf{x}'=\mathbf{x}/\xi$, (where $\xi$ defines the healing length of the system), $t'=t/\tau$, (where $\tau = \hbar/\mu$), and $\psi'=\psi/\sqrt{\mu/ g}$, where $\mu$ and $g$ are the chemical potential and interaction strength of the system. For brevity, primes will now be dropped.

Similar to 2D and 3D, the 4D GPE supports quantum vortex solutions; in the simplest case, a 4D vortex is described by a wave function of the form $\psi_1 (\mathbf{x}) = f_k(r_1) e^{i k \phi_1} $, where $(r_1, \phi_1)$ are the polar coordinates defined within a given 2D plane (e.g. the $x\!-\!y$ plane); the wave function's magnitude, $|\psi_1 (\mathbf{x}) |= f_k (r_1) $, is a real-valued function which vanishes as $r_1 \rightarrow 0$, corresponding to a vortex spanning the entire orthogonal 2D plane (e.g. the $z\!-\!w$ plane). 
As in lower dimensions, the phase of the field then winds around the vortex an integer $k$ number of times, inducing an in-plane superfluid velocity field ${\bf v}_1 = k {\hat{\bm \phi}}_1/r_1$. 

Hereafter we focus on the case $k=1$, which can be energetically stabilised by rotating the 4D superfluid with respect to the given 2D plane~\cite{mccanna2021superfluid} and we are interested in how two such vortices dynamically reconnect within a 4D superfluid. 
We define a reconnection of superfluid vortices in arbitrary dimensions as a point in time where the topology of the vortices changes. 
This is the standard definition used in 3D both for superfluids~\cite{Schwarz1985, Koplik1993, Alamri2008}, and -- originally -- for classical fluids where tubes of vorticity replace quantized vortices \cite{Kida1994}. 
More precisely, let \(\mathcal{C}(t) = \{\mathbf{x}\in \region \mid \psi(\mathbf{x},t) = 0\}\) be the set of vortices at time \(t\), where \(\region\subset \mathbb{R}^d\) is the region containing the superfluid (excluding its boundary): a reconnection then occurs at a time \(t=t^*\) if and only if \(\mathcal{C}(t^*)\) is not homeomorphic to \(\mathcal{C}(t)\) for times immediately before \(t^*\) or  immediately after (or both). Note that this definition assumes that \(\psi\) can only vanish on the boundary of the superfluid or on a vortex. This is not true in general, but it applies in all of our numerical simulations.

To better understand this definition of reconnection, we will first examine how it works for reconnections in 3D. Perhaps the simplest kind of 3D reconnection is that of two extended vortex lines. These are initially separated, then approach each other in an antiparallel fashion and touch at time \(t^*\), before separating again in a different configuration. In terms of the definition we give above, we can say that \(\mathcal{C}(t)\) in this case is not \emph{connected} for times other than \(t^*\), since it is a disjoint union of two lines. At \(t^*\) these lines meet and \(\mathcal{C}(t)\) becomes a connected space. Therefore its topology changes and we say that a reconnection occurs.

In 4D, the simple case we will consider is that of two initially intersecting skew vortex planes. In this case we will find \(\mathcal{C}(t)\) remains a connected space throughout the dynamics, but (except for special cases) it undergoes a topology change from \emph{simply-connected} (intersecting planes) to \emph{multiply-connected} (curved surface with a hole). We will discuss this in more detail later; firstly we will briefly review our previous results relating to the geometry and interactions of vortex planes in 4D. 

Similar to in lower dimensions, the dominant contribution to the interaction energy between two vortex planes in a large, homogeneous system is given by the hydrodynamic kinetic energy, 

\begin{equation}
    E_{kin} \approx \int  \frac{\rho}{2} {\bf v}^2 d^4 x,
    \label{eq:kinetic}
\end{equation}

where $\rho  = |\psi|^2$ is the density, and ${\bf v} = {\bf v}_1 + {\bf v}_2$ is the total velocity field found by summing the velocities of the individual vortices~\cite{mccanna2023curved}. 
This can therefore be decomposed into a sum of the energies of each individual vortex along with the hydrodynamic vortex-vortex interaction energy~\cite{mccanna2023curved}

\begin{equation}
    E_{vv} \approx \int  \rho {\bf v}_1 \cdot {\bf v}_2 d^4 x \label{eq:vv},
\end{equation}

which can be positive (i.e. repulsive) or negative (i.e. attractive), depending on the relative orientation of the two vortex planes.

We will parametrise these non-orthogonal vortex plane states symmetrically as follows: given \(\mathbf{x}\) the spatial coordinates in the lab frame, we define two new frames $ \pm $, with respective coordinates \(\mathbf{x}^\pm\). These are defined in relation to each vortex plane and are each tilted from the lab frame in an equal and opposite way to each other. 
That is, these vortex frames are related to the lab frame by \(\mathbf{x}^{\pm} = R(\pm\theta_1/2,\pm\theta_2/2)\mathbf{x}\), and therefore to each other by \(\mathbf{x}^+ = R(\theta_1,\theta_2)\mathbf{x^-}\), where \(R(\theta_1,\theta_2)\) is a double rotation which we can assume without loss of generality to take the form~\cite{mccanna2023curved}

\begin{equation}
     R(\theta_1,\theta_2) = 
    \begin{pmatrix}
        \cos{\theta_1} & 0 & -\sin{\theta_1} & 0 \\
        0 & \cos{\theta_2} & 0 & -\sin{\theta_2}  \\
        \sin{\theta_1} & 0 & \cos{\theta_1} & 0  \\
        0 & \sin{\theta_2} & 0 & \cos{\theta_2}
    \end{pmatrix},
    \label{eq:skewMatrix}
\end{equation}

where $\theta_{1,2} \in (-\pi/2, \pi/2)$.
We then have that one vortex lies in the \(\n{z}\dash\n{w}\) plane (inducing rotation in the \(\n{x}\dash\n{y}\) plane as described above), while the other lies in the $\p{x}\dash\p{y}$ plane (inducing rotation in the \(\p{z}\dash\p{w}\) plane).

The two angles $\theta_1$ and $\theta_2$ therefore define the relative initial orientation of the two vortices, which intersect at the origin. Note that the simplest case, $\theta_1\!=\!\theta_2\!=\!0$, corresponds to having two completely orthogonal intersecting vortex planes; in this case, $E_{vv}$ vanishes as the resulting velocity fields are also orthogonal. Unlike in lower dimensions, this intersecting vortex configuration can be energetically stabilised by a suitable ``double rotation" of the superfluid~\cite{mccanna2021superfluid} with two equal frequencies. More generally, if \(\theta_1\!=\!\theta_2\) then the two vortices are in an aligning configuration and can be energetically stabilised by a double rotation with generically unequal frequencies~\cite{mccanna2023curved}.

In this paper, we will focus on the dynamics associated with three classes of configurations:
the (1) \textit{aligning} case, when $\theta_1\!=\!\theta_2$, where the two vortices hydrodynamically repel each other, (2) the \textit{planar-aligning} case, when $\theta_2\!=\!0$ and $\theta_1\!>\!0$, where the hydrodynamic interaction energy vanishes by symmetry, and (3) an \textit{anti-aligning case}, when $\theta_1\!=\!-\theta_2$, where vortices hydrodynamically attract each other~\cite{mccanna2023curved}.
All the following angles are therefore  expressed in terms of $\theta$, where $\theta\!\equiv\!\theta_1$.

For short times \(t\) and small distances (i.e., $t$, $x<1$) from the reconnection point we can analytically examine the dynamics using the linearised GPE model (i.e. the Schr{\"o}dinger equation). This predicts that the evolution of the non-orthogonal planes is approximated by~\cite{mccanna2023curved2}
\begin{align}
    \psi = i(\sin\theta_1-\sin\theta_2)t + (\n{x} + i \n{y})(\p{z} + i \p{w}).
\end{align}
In particular, this predicts that aligning planes (\(\theta_1=\theta_2\)) do not reconnect -- and in fact form a stationary state -- which remains to be seen in the full nonlinear dynamics. 
In the general \(\theta_1 \neq \theta_2\) case the linearized dynamics predict a reconnection that always consists of a hole opening in the vortex surface~\cite{mccanna2023curved2}. The resulting surface is homeomorphic to a punctured plane - the interested reader can find a proof of this in Appendix A.
For that reason we only look at the special cases of aligning, planar-aligning, and anti-aligning tilt in this paper.

\subsection{ Numerical Simulations} We solve the 4D GPE (\ref{eq:4DGPE}) using a fourth-order Runge-Kutta scheme for time evolution and standard finite-difference methods for computing spatial derivatives. 
The spatial domain is discretised by a uniform step $\mathrm{dx}\!=\!0.5$ in each direction, which ensures that the vortex core ($\sim5\xi$) is well sampled numerically, and the time step is set to $\mathrm{dt}\!=\!0.025$. 
In all simulations, we include a hard-wall hyper-spherical potential with $V({\mathbf{x}})\! =\! 10$ if $r^2\!\geq\!R^2$, and which vanishes otherwise, where $r\!=\!\sqrt{x^2+y^2+z^2+w^2}$ and $R=30$ being the radius of the boundary. The wave function at $t\!=\!0$ is given by
\begin{equation}
    \psi(\mathbf{x}) = \psi_0(\mathbf{x}) F(\n{r}_1) e^{i \n{\phi}_1} F(\p{r}_2)  e^{i \p{\phi}_2} \label{eq:input}
\end{equation}
where \(\psi_0\) is the numerical solution for the ground state in the external potential with no vortices, (\(\n{r}_1,\n{\phi}_1\)) and (\(\p{r}_2,\p{\phi}_2\)) are polar coordinates in the ${\n{x}\dash\n{y}}$ and $\p{z}\dash\p{w}$ planes, respectively, and $F(r)$ is the radial function given by the Pad\'{e} approximation~\cite{berloff2004pade} of a vortex core on a uniform background. 
The results are ran briefly in imaginary time, before being run in real time to study dynamics. Note that visualisation of 4D vortices has obvious difficulties; we choose to depict the vortex surfaces in the $(x,y,z)$ subspace, with the $w$ direction denoted by the colour of the image, see e.g. Fig~\ref{fig:Fig_1}. 

\section{Numerical Results} \label{sec:results}
\subsection{Aligning Vortices} 
When $\theta_1\!=\!\theta_2\!=\!\theta$, the vortices are initially in an aligning orientation. For example, as $\theta\!\rightarrow\! \pi/2$, we see from Eq.~\ref{eq:skewMatrix} that \(\n{x} + i\n{y}\) and \(\p{z} + i\p{w}\) approach each other, meaning that the two vortices overlap in the same plane. More generally, for aligning vortices, it can be shown that Eq.~\ref{eq:vv} becomes
$    E_{vv} \!=\! -4\mu N \frac{\xi^2}{R^2}\ln(\cos{\theta})$,
where $\mu$ is the chemical potential and $N$ is the number of particles, corresponding to a positive interaction energy~\cite{mccanna2023curved}. Interestingly, we find that, despite this repulsive vortex-vortex interaction, aligned vortices do not reconnect in the real-time dynamics, but instead continue to intersect at the origin while rotating around each other freely, akin to the Abrikosov lattice in two- and three-dimensional systems \cite{abo2001observation,PhysRevLett.91.100402}. The lack of reconnection is consistent with previous results for a simple linear expansion of the short-term dynamics near the vortices~\cite{mccanna2023curved2}. Similar stationary states with intersecting vortex lines have also been found in 3D~\cite{meichle} (the simplest being two intersecting, orthogonal lines), although these require more symmetry than our aligning states, and cannot be energetically stabilised by rotation, unlike non-orthogonal vortex planes in 4D~\cite{mccanna2023curved}.

\begin{figure}
    \centering
    {\includegraphics[trim={1cm 10cm 2.5cm 10cm},clip,width=0.5\textwidth]{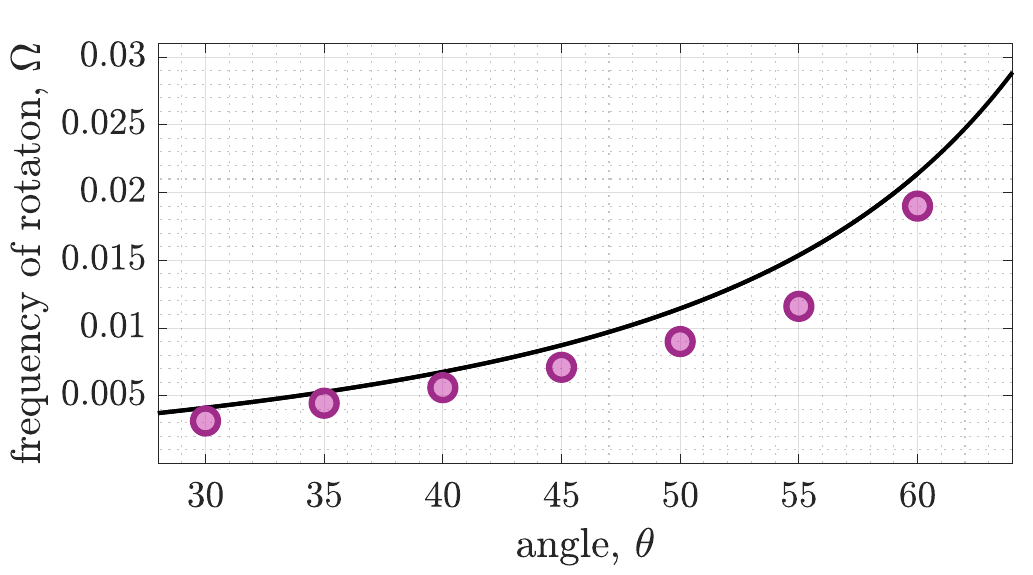}}
    \caption{The frequency of precession in the \(x\dash w\) plane, $\Omega_{xw}$, of two aligning vortices for each aligning angle, $\theta$, in a smaller system of radius \(R=18\). The analytical solution (\textit{black line}) from Eq.~\eqref{eq:rotation} is compared to the frequency of the rotation measured computationally (\textit{purple}).}
\label{fig:rotation}
\end{figure}

As we now derive, we can find an analytical expression for the precession frequency as a function of the tilt angle of two aligning vortices by assuming that the motion is a rigid (double) rotation of the entire superfluid described by an angular frequency tensor \(\mathbf{\Omega}\). In other words, this means that the aligning vortex planes form a stationary state in a reference frame rotating according to \(\mathbf{\Omega}\). The energy \(E'[\psi]\) in this rotating frame is related to that of the lab frame \(E[\psi]\) by the usual formula \(E'[\psi] = E[\psi] - \mathbf{\Omega}\cdot\mathbf{L}[\psi]\). Then, for any tilt angle \(\theta\), the aligning state \(\psi_\theta\) will only be stationary if it extremises the energy \(E'\), such that \(\partial E'[\psi_\theta] / \partial \theta = 0\). This then gives us the following equation for the angular frequency tensor of the precession
\begin{align}
    \mathbf{\Omega}\cdot\frac{\partial \mathbf{L}[\psi_\theta]}{\partial\theta} = \frac{\partial E[\psi_\theta]}{\partial\theta}.
    \label{eq:AngFreq}
\end{align}
As derived in Appendix B, the angular momentum of aligning skew vortex plane states can be written as 
\begin{align}
    \mathbf{L}[\psi_\theta] = N\left( \hatbf{x}\wedge\hatbf{y} + \hatbf{z}\wedge\hatbf{w} \right) + N\sin\theta\left( \hatbf{x}\wedge\hatbf{w} + \hatbf{z}\wedge\hatbf{y} \right),
\end{align}
which can then be used to calculate the LHS of Eq.~\ref{eq:AngFreq}. To calculate the RHS, we use the usual formula for  total energy in the inertial lab frame, \(E[\psi] = \frac{1}{2}\int \left(|\nabla\psi|^2 + |\psi|^4\right) \diff^4r\), expressing this hydrodynamically in terms of \(\rho\) and the velocity field from each vortex \(\mathbf{v}_{1,2}\) as
\begin{align}
    E[\psi] &= \int \left(\frac{1}{2}\rho\mathbf{v}_1^2 + \frac{1}{2}\rho\mathbf{v}_2^2 + \rho\mathbf{v}_1\cdot\mathbf{v}_2 + \frac{1}{2}|\nabla\sqrt{\rho}|^2 + \frac{1}{2}\rho^2 \right) \diff^4 x,
\end{align}
where the first two terms are the kinetic energy of each vortex plane separately, the third is the hydrodynamic interaction between them, the fourth is the quantum pressure energy, and the last term is the interparticle interaction energy. If we evaluate this energy for an aligning state, \(\psi_\theta\), then the first two terms do not depend on \(\theta\) due to the spherical symmetry of the system, and we will assume that the variation of last two terms with \(\theta\) is negligible. This leaves us with the hydrodynamic interaction as the only term that significantly varies with theta, which for aligning states is approximated (assuming constant density \(\rho\)) by~\cite{mccanna2023curved} 
\begin{align}
    E_{vv}[\psi_\theta] = -4\mu N \frac{\xi^2}{R^2}\ln\cos\theta.
\end{align}
Using this in Eq~\eqref{eq:AngFreq} gives the following equation for the precession frequencies
\begin{align}
    \Omega_{xw} + \Omega_{zy} = 4\mu\frac{\xi^2}{R^2}\frac{\tan\theta}{\cos\theta} .
\end{align}
Assuming, by symmetry, that \(\Omega_{xw} = \Omega_{zy} = \Omega\) we then get the following equation for the frequency of precession in dimensionless form
\begin{align}
    \Omega = \frac{2}{R^2}\frac{\tan\theta}{\cos\theta},
    \label{eq:rotation}
\end{align}
which agrees with the computational results, as shown in Fig. \ref{fig:rotation}.

\begin{figure}
    \centering
{\includegraphics[trim={3cm 3cm 4cm 5.5cm},clip,width=0.45\textwidth]{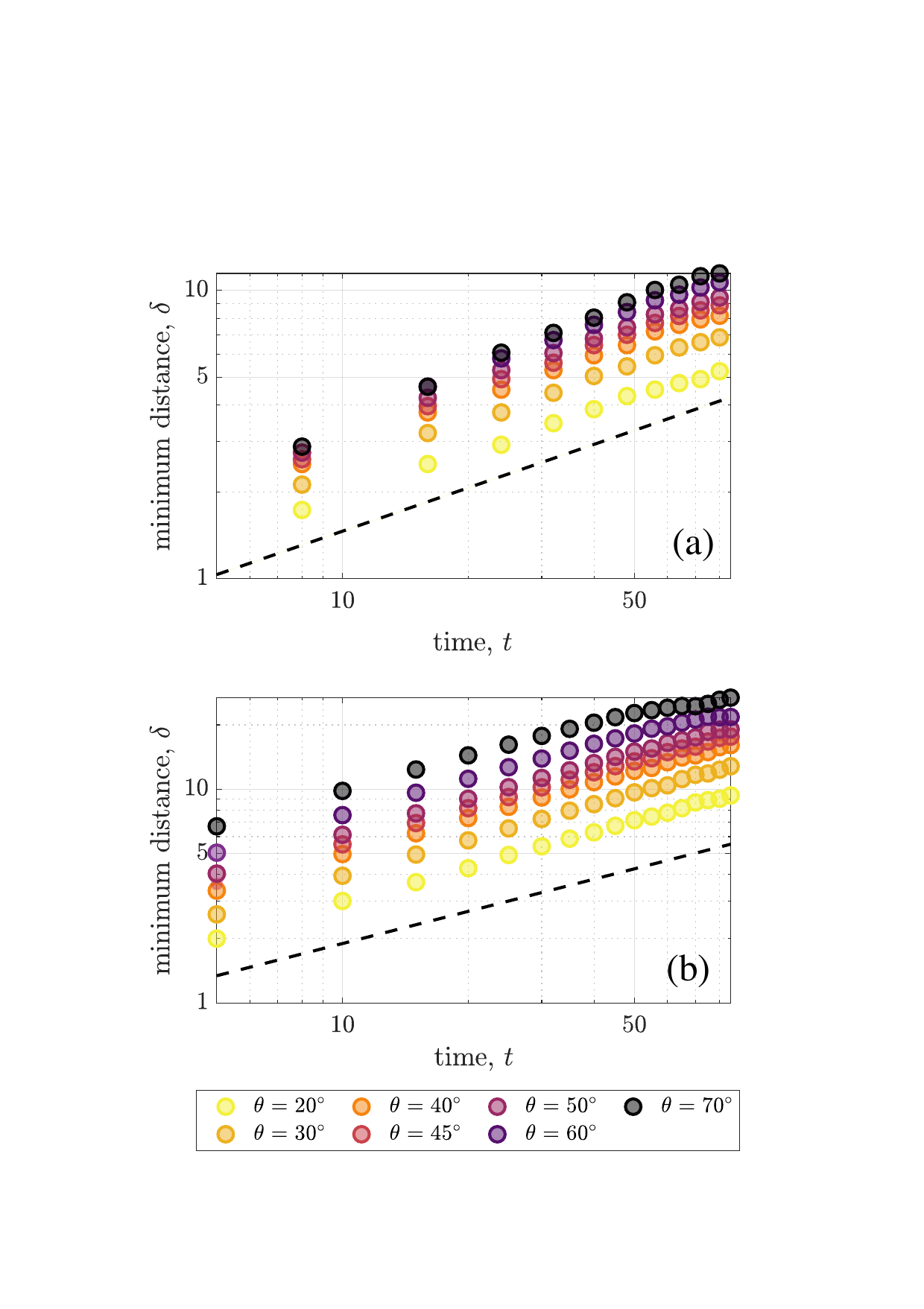}}
    \caption{The change in the minimum distance, $\delta$, between vortices (as defined in the main text)  within the reconnection zone as obtained from the real-time dynamics [c.f. Fig~\ref{fig:Fig_1}]  for (a) initially planar aligned vortices and (b) initially anti-aligned vortices. Both orientations show a post-reconnection scaling law, $\delta (t) \sim t^{1/2}$, matching the 3D result in Eq. (\ref{eq:reconnection}). The scaling $t^{1/2}$ (\textit{dashed}) is marked for reference.} 
    \label{fig:reconnection_min_dist}
\end{figure}

\subsection{ Planar-aligning Vortices} In this case, we have that $\theta_1\!=\!\theta\!>\!0$ and $\theta_2\!=\!0$, corresponding to the \(\n{x}\) and \(\p{z}\) axes being tilted towards each other, while \(\n{y} \!=\! y\) and \(\p{w} \!=\!w\) are unaffected. In this special case, the hydrodynamic interaction between the vortices vanishes by symmetry, regardless of the value of \(\theta\)~\cite{mccanna2023curved}. When $\theta\rightarrow \pi/2$, both \(\n{x}\) and \(\p{z}\) approach \((x+z)/\sqrt{2}\), such that the two vortices intersect along the line given by \(x = -z\), \(y = w = 0\). In any 3D cross-section of the system orthogonal to this line the vortices will appear as two intersecting lines. We can therefore expect this limit to give a 4D generalisation of the usual 3D vortex reconnection physics~\cite{nazarenko2003analytical}.

Numerically, we find that in our real-time dynamics, two planar-aligning vortices will indeed reconnect, causing a hole to open up at the intersection point around the origin, as shown in Fig.~\ref{fig:Fig_1}(a,b,c) for $\theta=50^{\circ}$. Unlike for extended vortex lines in 3D, the vortex remains a single connected object throughout the real-time dynamics. Instabilities along the vortex surface also form during the reconnection, shown as small density ripples along the vortex at large post-reconnection times.

To quantify the reconnection dynamics, we describe the hole around the origin as an ellipse and extract the semi-minor axis as a measure of the minimum distance between the vortices arising from the reconnection. This approximation is found to give a precise description of the size of the reconnection region and is similar to considering the minimum distance between two vortex lines after they reconnect in 3D~\cite{paoletti2010reconnection,zuccher2012quantum,allen2014,villois2017universal,galantucci2019crossover,PhysRevFluids.5.104701}. By measuring this minimum distance within the reconnected region over time, we find that the same universal scaling
law in Eq.~(\ref{eq:reconnection}) also holds for reconnection of two planar-aligning vortices in four dimensions, as can be seen in Fig.~\ref{fig:reconnection_min_dist}(a). 
This aforementioned scaling corresponds to $\sim t^{1/2}$ (Eq. \ref{eq:reconnection}), which is plotted for reference in Fig.~\ref{fig:reconnection_min_dist}(a), and verifies that the scaling is observed in the planar reconnection case, for all reconnection angles $\theta$. 

To further analyse the reconnection dynamics, we can also monitor the incompressible energy, $E_i$ - i.e. the energy associated with the vortex core - as shown in Fig.~\ref{fig:incompressible_energy} (a). The typical way of calculating the incompressible energy is through a Helmholtz decomposition; however, this is not valid in a 4D system, and thus a generalised Helmholtz decomposition is used instead. In more detail, given a vector field $ {\bf F}({\bf r}) $ with $ {\bf r} \in \mathbb{R}^d $, we perform the Hodge-Helmholtz decomposition
\begin{equation} 
{\bf F} = -\nabla\Phi + {\bf R}
\label{eq:hhd}
\end{equation}
with $ \nabla $ the gradient operator in $ d $ dimensions, $ \Phi({\bf r}) $ and $ {\bf R}({\bf r)} $ being the scalar potential and the solenoidal (or rotational) field, respectively, by rewriting the identity in Fourier space.
Expressing the Fourier transforms as
\begin{equation} 
G_{(\cdot)}({\bf k}) = \frac{1}{(2\pi)^d} \int (\cdot)({\bf x}) e^{-i {\bf k}\cdot{\bf r}} \diff^4 x
\end{equation}
and
\begin{equation}
(\cdot)({\bf r}) = \int G_{(\cdot)}({\bf k}) e^{i {\bf k}\cdot{\bf r}} \diff^4 x
\end{equation}
the decomposition [Eq~\eqref{eq:hhd}] in Fourier space gives
\begin{equation} 
{\bf G}_{\bf F} = -i{\bf k}G_\Phi + {\bf G}_{\bf R} \, .
\end{equation}
Taking the scalar product of the last expression with $ i{\bf k} $ and remembering that the orthogonality between the gradient of the scalar field and solenoidal field yields
\begin{equation} 
i {\bf k}\cdot {\bf G}_{\bf F} = |{\bf k}|^2 G_\Phi
\quad \Longrightarrow \quad
G_\Phi = i \frac{{\bf k}\cdot {\bf G}_{\bf F}}{|{\bf k}|^2} \, ,
\end{equation}
hence
\begin{equation}
G_{\bf R}= {\bf G}_{\bf F} - \frac{{\bf k}\cdot {\bf G}_{\bf F}}{|{\bf k}|^2} {\bf k} \, .
\end{equation}
Finally, by transforming back to physical space one obtains
\begin{equation} 
\Phi({\bf r}) = \int {G}_\Phi e^{i {\bf k}\cdot{\bf r}} \diff^4 k 
\end{equation}
and
\begin{equation} 
{\bf R} = \int {\bf G}_{\bf R} e^{i {\bf k}\cdot{\bf r}} \diff^4 k \, .
\end{equation}
To use this generalised Hodge-Helmholtz decomposition to calculate the incompressible energy, we start from the hydrodynamic kinetic energy Eq. \eqref{eq:kinetic}, which we rewrite as
\begin{equation}
    E_{kin} = \int |\mathbf{f}|^2 \diff^4 x 
\end{equation}
by defining the following vector field
\begin{equation}
{\bf f} = \frac{1}{\sqrt{2}}\sqrt{\rho}{\bf v}.
\end{equation}
We can decompose this vector field as \(\mathbf{f} = \mathbf{f}_c + \mathbf{f}_i\) into its compressible \(\mathbf{f}_c\) (gradient of scalar potential) and incompressible \(\mathbf{f}_i\) (solenoidal) parts using the Hodge-Helmholtz decomposition, such that
\begin{equation}
|{\mathbf f}|^2 = |{\bf f}_c|^2 + |{\bf f}_i|^2 \, .
\end{equation}
Therefore, the kinetic energy also follows the decomposition, namely
\begin{equation}
E_{kin} = E_{c} + E_{i} \, ,
\end{equation}
with 
\begin{equation}
E_c = \int |{\bf f}_c|^2 \diff^4 x 
\end{equation}
and 
\begin{equation}
E_{i} = \int |{\bf f}_i|^2 \diff^4 x \, 
\end{equation}
being the compressible and incompressible kinetic energies respectively.

By monitoring the incompressible kinetic energy for multiple reconnections with different $\theta$ values, we observe that the reconnection process for planar-aligning vortices is irreversible, with the energy associated with the vortex core being transferred into sound energy. There is also formation of Kelvin waves along the vortex plane. This energetic transfer is shown in Fig. \ref{fig:incompressible_energy}(a), where we observe a clear downwards trend in the incompressible energy over time, becoming more marked with increasing values of $\theta$ as the initial vortex configuration tends towards the 3D-like limit. We have therefore shown that the reconnection of two planar vortices in a 4D condensate follows all of the generalised laws observed in the reconnection of vortices in a 3D condensate.

\subsection{Anti-aligning Vortices} 

We lastly study the case of anti-aligning vortices corresponding to $\theta = \theta_1 \!=\!-\theta_2$. Then as $\theta\!\rightarrow\! \pi/2$, we have that \(\n{x} + i\n{y}\) and \(\p{z} + i\p{w}\) become complex conjugates of each other, meaning that the two vortices anti-align within the same plane (and would potentially annihilate). More generally, it can be shown that the hydrodynamic vortex-vortex interaction energy has the same form as stated for the aligning case above, except with the opposite sign, corresponding to a negative interaction energy~\cite{mccanna2023curved}.

\begin{figure}
    \centering
    {\includegraphics[trim={2cm 3cm 2cm 4cm},clip,width=0.5\textwidth]{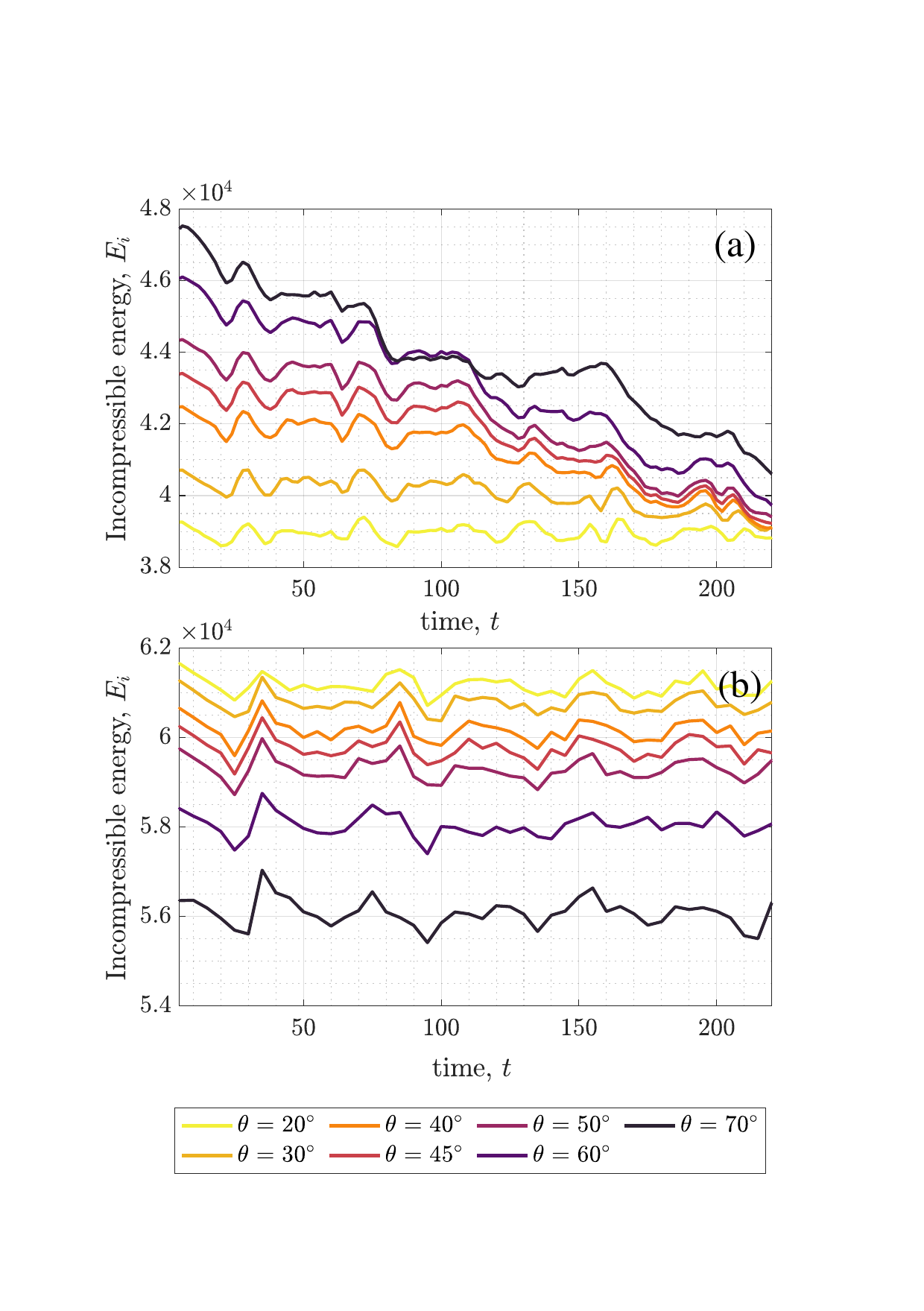}}
    \caption{The incompressible energy, $E_i$, over time as calculated from the numerical real-time dynamics, after the initial reconnection at $t=0$ for different alignment angles for the (a) planar orientation and the (b) anti-aligning orientation. In the planar reconnection case, (a), we see that the incompressible energy falls for all values of $\theta$, matching known 3D results. The anti-aligning case, however, shows a lack of loss of incompressible energy, signalling a lack of the significant sound wave emission typically observed in 3D reconnection.}
\label{fig:incompressible_energy}
\end{figure}

Numerically, our real-time dynamical simulations find that two anti-aligned vortices will reconnect around the origin, as shown in Fig. \ref{fig:reconnection_min_dist}(b) for $\theta=50^{\circ}$. This is similar to the planar-aligning case and is also consistent with a previous prediction from a simple linear argument for the short-time dynamics near the vortices~\cite{mccanna2023curved2}, like in the aligning case. We can also again quantitatively characterise the 4D reconnection dynamics by plotting the minimum distance obtained from the semi-minor axis of the reconnection zone in Fig. \ref{fig:reconnection_min_dist}(b). We again see a clear $t^{1/2}$ scaling, in agreement with the scaling of both 3D reconnection dynamics and the 4D planar-aligned case discussed above, with a small flattening out at the late times due to finite size effects. However, an interesting and immediate difference here is the lack of fluctuations of the vortex planes. The incompressible energy also does not significantly vary on average over time, as shown in Fig.~\ref{fig:incompressible_energy} (b), suggesting that the reconnection may be reversible.

\section{Discussion and Conclusions}
\label{sec:concl}


{This work provides the first systematic study of vortex dynamics in a condensate with an extra physical dimension. We have characterised three different ways that quantum vortices can interact and reconnect in a 4D superfluid, going beyond the physics of lower-dimensional systems.} Firstly, when the two vortices are in an aligning orientation, there is no reconnection of the intersection point and instead the vortices rotate around each other in real time. Secondly, we showed that, when in a planar-aligning orientation, the reconnection dynamics does resemble that in 3D, with the same scaling behaviour and an irreversible transfer of incompressible energy. Finally, we discovered a qualitatively new reconnection regime emerging in 4D when the two vortices are anti-aligned, in which the average incompressible energy does not significantly vary implying the reconnection mechanism may be time reversible. 

This unusual behaviour is markedly different from lower dimensions, reflecting the rich phenomenology that materialises in 4D superfluids. 
Going further, there are multiple interesting avenues to explore. Firstly, it would be important to explore the initiations of the instabilities in the post-reconnection state in the planar reconnection, and the characterisation of the Kelvin waves formed. Another avenue would be in understanding the consequences of a lack of loss of incompressible energy during the reconnection, and how this changes long term multi-vortex dynamics, such as turbulence. An important open question is whether the dynamics in the anti-aligning case is in fact reversible or not; this could be studied e.g. through examining the evolution of the incompressible energy spectrum in a large enough system, or by finding an initial state which will evolve towards a reconnection, allowing us to examine in more detail how the reconnection occurs. 
It would also be interesting to extend our study to non-orthogonal vortex pairs with higher winding numbers $k$, which are expected to be dynamically-unstable towards separation into many vortex planes that could then interact in complex ways. In the very large $k$ limit, this may lead to the formation of Abrikosov-lattice-like structures with multiple reconnection regions. In the future, it will also be interesting to identify and study closed vortex surfaces, which do not touch the system boundary, such as the analogue of 3D vortex rings and knots~\cite{Proment2012,Proment2014,Scheeler2014,villois2017universal,villois2020irreversible,PhysRevFluids.5.104701}, and other more exotic types of closed surfaces that may be possible in 4D~\cite{gallier2013}. Our work may also be extended to more complex 4D order parameters, motivated by the appearance of non-Abelian vortices~\cite{Kawaguchi,Machon} in 3D spinor condensates, which exhibit richer intersection and reconnection physics. 

Finally, our work can be extended to more realistic experimental settings, based on techniques such as synthetic dimensions. For example, most current approaches to engineering a synthetic dimension result in a discretized dimension~\cite{Boada2012,Celi2014,Stuhl2015,Mancini2015,Gadway2015, Livi2016,Ozawa2016,Yuan2016,Kolkowitz2017,Martin2017, sundar2018synthetic, Price2017, chalopin2020exploring,Kanungo2022, lustig2019photonic, Dutt2020,cai2019experimental, price2019synthetic,boyers2020exploring,lienhard2020realization, oliver2021bloch, oliver_photon, cornish2024quantum, ehrhardt2023perspective, ozawa2019topological, senanian2023programmable}. Depending on the size of the typical vortex length-scales as compared to this discrete lattice, a future experiment could either be in a continuum limit, similar to that studied here, or in a regime where discretization becomes important, which will enrich the physics and require further study. We also have assumed in our numerical simulations that the bosonic quantum fluid is contained within a hard-wall hyper-spherical boundary, which preserves rotational symmetry. However, in practice, differences between real and synthetic dimensions will likely break this symmetry, imposing different boundary conditions on the condensate and revising the energetic arguments considered above. In a similar vein, synthetic-dimension experiments can have many particular features, which depend on the specific implementation and which can include nonuniform hoppings, nonequilibrium effects from external driving and long-range interactions along the synthetic dimension~\cite{Celi2014, ozawa2019topological}. In the future, it will be interesting to include some or all of these effects to  model the interactions and reconnections of topological phase defects in complex geometries using different combinations of real and synthetic dimensions.\\

\begin{acknowledgments}
We thank Mike Gunn and Mark Dennis for helpful discussions. 
This work is supported by the Royal Society via grants
UF160112, URF\textbackslash R\textbackslash221004, RGF\textbackslash{}EA\textbackslash{}180121 and RGF\textbackslash{}R1\textbackslash{}180071 and by the Engineering and Physical Sciences Research Council [grant numbers EP/W016141/1 and EP/Y01510X/1].
\end{acknowledgments}

\begin{appendices}

\section*{Appendix A: Topology of reconnecting vortex planes}
\label{app:homeomorphism}
Here we will show that the vortex core associated with a pair of skew vortex planes after a reconnection is homeomorphic (i.e. topologically equivalent) to a single punctured plane. This formalizes the observation that the reconnected surface contains a hole, as can be observed in Fig~\ref{fig:Fig_1}. The change in topology from a simply connected space (a pair of intersecting planes) to a multiply connected one shows that the dynamics is truly a reconnection process in the usual sense. Firstly, two topological spaces \(A\) and \(B\) are homeomorphic if there exists a continuous bijective function between them \(h: A \to B\) whose inverse \(h^{-1}: B \to A\) is also continuous. This function \(h\) is called a homeomorphism between \(A\) and \(B\). Now we will restate the linearised description of the reconnecting vortex core [Eq(5) in the main text], which is given by the following set
\begin{align}
    V = \{\mathbf{x}\in\mathbb{R}^4 \mid (\n{x} + i\n{y})(\p{z} + i\p{w}) = i\gamma\},
\end{align}
where \(\gamma = t(\sin\theta_2 - \sin\theta_1)\) and all relevant quantities are as defined in the main text. Technically we need to specify a topology on \(V\) before we can talk about continuity of functions. As usual for subspaces we will use the ``subspace topology"~\cite{Munkres2000Topology} which is the one induced by the Euclidean topology in four dimensions (the standard topology for \(\rLine^4\)). This means that a function \(f: V \to A\) for any codomain \(A\) is continuous if it is the restriction of a continuous function \(g: \mathbb{R}^4 \to A\) under the Euclidean topology. Next, without loss of generality, we will fix a value of \(\gamma > 0\) (i.e. \(\theta_1 > \theta_2\) and \(t\) fixed), which will allow us to remove \(\gamma\) by a rescaling of coordinates. We will also ``unskew" the vortex core using a linear transformation. Combining both of these, we have the following linear function
\begin{align}
    F(\mathbf{x}) =
    \sqrt{\gamma}
    \begin{pmatrix}
        \mathcal C & \mathcal{S} \\
        \mathcal{S} & \mathcal{C}
    \end{pmatrix}^{-1} \mathbf{x},
\end{align}
where \(\mathcal{C} = \diag(c_1, c_2)\), and \(\mathcal{S} = \diag(s_1, s_2)\), and we have used the shorthand \(c_j = \cos(\theta_j/2)\), \(s_j = \sin(\theta_j/2)\). This function is defined provided that the given matrix inverse exists, which is true provided \(\cos\theta_1\cos\theta_2 \neq 0\). We are only considering tilt angles in the range \(\theta_j \in (-\pi/2,\pi/2)\), so this is automatically satisfied. Note that the excluded cases correspond to \(\n{x} = \p{z}\) when \(\theta_1 = \pm\pi/2\), or \(\n{y} = \p{w}\) when \(\theta_2 = \pm \pi/2\). In these cases the initial vortex planes intersect along a line, rather than a point, and according to the linearised dynamics they reconnect in the same way that vortex lines in 3D do --- they form two disconnected surfaces after the reconnection. For the general case the function is an invertible linear map, and so it is a continuous bijection with a continuous inverse, i.e. a homeomorphism, from \(V\) to the space
\begin{align}
    F(V) = \{\mathbf{x}\in\mathbb{R}^4 \mid (x + iy)(z + iw) = i\}.
\end{align}

This is equivalent to a hyperbola over complex variables, which we define as \(\cHyp = \{(\zeta_1,\zeta_2) \in \cPlane^2 \mid \zeta_1\zeta_2 = 1 \} \). To show that \(\cHyp\) is homeomorphic to the punctured plane given by \( \pPlane = \mathbb{C} - \{0\} \), we will use a very simple homeomorphism given by projection onto either the first or second complex coordinate (i.e. either of \(\zeta_{1,2}\)). We will arbitrarily choose the first, such that our homeomorphism \(h: \cHyp \to \pPlane\) is defined as
\begin{align}
    h(\zeta_1,\zeta_2) = \zeta_1.
\end{align}
This is clearly injective (one-to-one) since \(h(\zeta_1,\zeta_2) = h(\zeta'_1,\zeta'_2) \iff \zeta_1=\zeta'_1\) from the definition of \(h\), and therefore \(\zeta_2 = 1/\zeta_1 = 1/\zeta'_1 = \zeta'_2\) from the definition of \(\cHyp\). It is also surjective (onto) since for any \(\zeta \in \pPlane\) we have the pair \((\zeta_1,\zeta_2) = (\zeta,1/\zeta) \in \cHyp\) such that \(h(\zeta_1,\zeta_2)=\zeta\). This means that \(h\) is indeed bijective, and that its inverse can be defined for all \(\zeta \in \pPlane\) as
\begin{align}
    h^{-1}(\zeta) = \left( \zeta, 1/\zeta \right).
\end{align}

All that's left is to show that both \(h\) and \(h^{-1}\) are continuous. Here we will use several well known results~\cite{Munkres2000Topology}. Firstly, for any topological product space \(A_1 \times A_2\) we can define projection functions, \(p_{i}: A_1 \times A_2 \to A_i\) for \(i\in\{1,2\}\), onto each factor. These take the form \(p_{i}(a_1,a_2) = a_i\) and are both continuous functions. Secondly, any continuous function with a domain \(A\) and range \(B\) can also be considered as a continuous function from \(X\) to \(Y\) provided that \(X\) is a subset of \(A\) and that \(Y\) contains \(f(X)\), the image of the restricted domain. Our function \(h\) can be defined as the projection function \(p_1: \cPlane^2 \to \cPlane\), with domain restricted to \(\cHyp\) and range restricted to \(\pPlane = p_1(\cHyp)\) and is therefore continuous. Similarly the function \(t: \pPlane \to \cPlane^2\) defined as \(t(\zeta) = (\zeta, 1/\zeta)\) is also continuous as both of its component functions are continuous~\cite{Munkres2000Topology}, and we can define \(h^{-1}\) by restricting the range of \(t\) to its image, \(\cHyp = t(\pPlane)\). Therefore \(h^{-1}\) is also continuous. This means that \(h\) is a homeomorphism between \(\cHyp\) and \(\pPlane\), and so these two spaces are homeomorphic.

\section*{Appendix B: Angular momentum of skew intersecting vortex planes}
\label{sec:AngMom}

In this Appendix, we calculate the angular momentum of a generic state containing two non-orthogonal intersecting vortex planes. As in the main text, we will let one vortex span the \(\n{z}\!-\!\n{w}\) plan, inducing circulation in the \(\n{x}\!-\!\n{y}\) plane, while the other spans the \(\p{x}\!-\!\p{y}\), inducing circulation in the \(\p{z}\!-\!\p{w}\) plane. The transformation between the various bases is given by Eq~(4) in the main text, which we will restate here in the following block matrix form
\begin{align}
    \mathbf{x}^{\pm} =
    \begin{pmatrix}
        \mathcal{C} & \mp\mathcal{S} \\
        \pm\mathcal{S} & \mathcal{C}
    \end{pmatrix}
    \mathbf{x},
    \label{eq:vortexCoords}
\end{align}
where \(\mathcal{C} = \diag(c_1, c_2)\), and \(\mathcal{S} = \diag(s_1, s_2)\), and we have used the shorthand \(c_j = \cos(\theta_j/2)\), \(s_j = \sin(\theta_j/2)\). 

We then start with the following ansatz for the order parameter describing the vortex pair
\begin{align}
    &\psi = \left( \n{x}+i\n{y} \right) \left( \p{z}+i\p{w}\right) g( \n{r}_1, \p{r}_2 ),
    \label{eq:ansatz}
\end{align}
where \(\n{r}_1\) and \(\p{r}_2\) are polar radii in the \(\n{x}\!-\!\n{y}\) and \(\p{z}\!-\!\p{w}\) planes, respectively, and \(g\) is a real-valued and positive-definite (but otherwise unspecified) function of these radii. First we will compute the angular momenta of this state, as given by the expectation of the usual operator, \(\hat{\mathbf{L}} = -i\mathbf{x} \wedge \nabla\). However, note that angular momentum cannot be transformed into a pseudovector in 4D as it can in 3D so we will need to treat it properly as an antisymmetric tensor. For our purposes, this means using the wedge product which generalises the 3D cross-product. Evaluating this operator on our ansatz using the product rule gives the following expression for the expectation value
\begin{align}
    \mathbf{L} = -i\int \mathbf{x} \wedge \biggl[ \frac{\nabla\left( \n{x}+i\n{y} \right)}{ \n{x}+i\n{y}} &+ \frac{\nabla \left( \p{z}+i\p{w}\right)}{\p{z}+i\p{w}} \nonumber \\ &+ \frac{\nabla g}{g} \biggr] \rho \diff^4 x,
\end{align}
where we have suppressed the arguments of \(g\) for brevity. The last term is pure imaginary, and so must vanish since angular momentum is a Hermitian operator. Next we will evaluate the derivatives and wedge products, using the \(\n{\mathbf{x}}\) coordinate frame for the first term and the \(\p{\mathbf{x}}\) frame for the second, which yields 
\begin{equation}
\begin{split}
    \mathbf{L} &=  \int \biggl[ \n{\hatbf{x}} \wedge \n{\hatbf{y}} + \p{\hatbf{z}} \wedge \p{\hatbf{w}} \\
    &\hphantom{=\int} - i
    (\n{z}\n{\hatbf{z}} + \n{w}\n{\hatbf{w}} ) \wedge \frac{ \n{\hatbf{x}} + i\n{\hatbf{y}} }{ \n{x} + i\n{y} } \\
    &\hphantom{=\int} - i (\p{x}\p{\hatbf{x}} + \p{y}\p{\hatbf{y}} ) \wedge \frac{ \p{\hatbf{z}} + i\p{\hatbf{w}} }{ \p{z} + i\p{w} } \biggr] \rho \diff^4x,
    \end{split}
\end{equation}
where \(\n{\hatbf{x}}\) denotes the unit vector in the \(\n{x}\) direction (and similarly for the other coordinates). The last two terms vanish by symmetry, as the integrands are odd functions of  \((\n{z}, \n{w})\), and \((\p{x}, \p{y})\), respectively. This then gives us the following simple result without any other assumptions on the density 
\begin{align}
    \mathbf{L} =  N \left( \n{\hatbf{x}} \wedge \n{\hatbf{y}} + \p{\hatbf{z}} \wedge \p{\hatbf{w}} \right).
\end{align}
Substituting in Eq~\eqref{eq:vortexCoords} we then get the following Cartesian components of \(\mathbf{L}\)
\begin{equation}
\begin{split}
\mathbf{L} &= Nc_- ( \hatbf{x} \wedge \hatbf{y} + \hatbf{z} \wedge \hatbf{w} ) +
    Ns_+ ( \hatbf{x} \wedge \hatbf{w} + \hatbf{z} \wedge \hatbf{y} ),
    \label{eq:AngMom}
\end{split}
\end{equation}
where \(\theta_\pm = (\theta_1 \pm \theta_2)/2\), \(c_- = \cos\theta_-\), and \(s_+ = \sin\theta_+\). For anti-aligning planes, we have that \(\theta_+ = 0\), and \(\theta_- = \theta \in (-\pi/2,\pi/2)\), such that the total angular momentum \(|\mathbf{L}|^2\) decreases as the magnitude of the tilt \(\theta\) increases, as expected. For planar-aligning vortices (\(\theta_+ = \theta_- = \theta\)) \(|\mathbf{L}|^2\) remains constant, while for aligning vortices (\(\theta_+ = \theta, \theta_- =0\)) \(|\mathbf{L}|^2\) increases with \(\theta\). Interestingly, aligning vortices have constant angular momenta in the original \(x\dash y\) and \(z\dash w\) planes, regardless of tilt. This can be explained using a special property of rotations in 4D known as isoclinic symmetry, as has been found previously~\cite{mccanna2023curved}.

\end{appendices}
\bibliography{apssamp.bib}

\end{document}

%% file: commands.tex
\newcommand{\diff}{\mathrm{d}}

\DeclareMathOperator{\diag}{diag}

\newcommand{\rLine}{\mathbb{R}}
\newcommand{\cPlane}{\mathbb{C}}
\newcommand{\pPlane}{\mathbb{C}^{\times}}
\newcommand{\cHyp}{H_\cPlane}
\newcommand{\region}{D}

\newcommand{\hatbf}[1]{\hat{\mathbf{#1}}}

\newcommand{\n}[1]{#1^-}
\newcommand{\p}[1]{#1^+}

\newcommand{\dash}{\textbf{--}\,}